\documentclass[12pt]{article}
 \usepackage{graphicx}
 \usepackage[cp1251]{inputenc}
 \usepackage{rotating}
\begin{document}
 \noindent {\footnotesize\it Astronomy Letters, 2004, Vol. 30, No. 4, pp. 251--257.}
 \newcommand{\dif}{\textrm{d}}

 \noindent
 \begin{tabular}{llllllllllllllllllllllllllllllllllllllllllllll}
 & & & & & & & & & & & & & & & & & & & & & & & & & & & & & & & & & & & & & \\\hline\hline
 \end{tabular}

  \vskip 0.5cm
  \centerline {\bf KINEMATIC CONTROL OF THE INERTIALITY OF ICRS CATALOGS}
  \bigskip
  \centerline {V.V.~Bobylev}
  \bigskip
  \centerline{\small\it Pulkovo Astronomical Observatory, St. Petersburg,  Russia}
  \bigskip
  \bigskip
{\bf Abstract}—We perform a kinematic analysis of the Hipparcos
and TRC proper motions of stars by using a linear
Ogorodnikov–Milne model. All of the distant ($r>0.2$~kpc) stars of
the Hipparcos catalog have been found to rotate around the
Galactic $y$ axis with an angular velocity of
$M_{13}^-=-0.36±0.09$~mas yr$^{-1}$. One of the causes of this
rotation may be an uncertainty in the lunisolar precession
constant adopted when constructing the ICRS. In this case, the
correction to the IAU (1976) lunisolar precession constant in
longitude is shown to be $\Delta p_1=-3.26±0.10$~mas yr$^{-1}$.
Based on the TRC catalog, we have determined the mean Oort
constants:
 $A=14.9\pm1.0$ and $B=-10.8\pm0.3$ km s$^{-1}$ kpc$^{-1}$.
The component of the model that describes the rotation of all TRC
stars around the Galactic $y$ axis is nonzero for all magnitudes,
$M_{13}^-=-0.86\pm0.11$~mas yr$^{-1}$.


\bigskip
\bigskip
\leftline {\hskip6mm INTRODUCTION}
\bigskip
Based on the Ogorodnikov–Milne model and using the proper motions
of stars, Clube (1972, 1973), du Mont (1977, 1978), and Miyamoto
and S\^oma (1993) modeled the Galactic rotation. This modeling
showed that, apart from the Galactic rotation parameters, it is
possible to control the inertiality of the catalogs being analyzed
and to refine the adopted precession constant. Here, instead of
the currently popular method for solving the Ogorodnikov–Milne
model equations using parallaxes, we use a method for determining
the kinematic parameters of the Galactic rotation that is
completely free from parallax errors to analyze Hipparcos data
(ESA 1997). In this method, the parallactic factor is assumed to
be equal to unity. First, this assumption makes it possible to
analyze stars even with negative parallaxes (the Hipparcos
catalog). Second, since the proper motions of TRC stars (H\o g et
al. 1998) can be analyzed in full only by this method, this
approach yields comparable data for both ICRS catalogs
(International Celestial Reference System). The kinematic method
for controlling the inertiality of the catalogs of stellar proper
motions is based on the analysis of the two components of the
rigid-rotation tensor that describe the rotation about the
Galactic y and x axes in the Galactic coordinate system.

\bigskip
\leftline {\hskip6mm BASIC EQUATIONS}
\bigskip
In this paper, we use a rectangular Galactic coordinate system
with the axes directed away from the observer toward the Galactic
center ($l=0^\circ$, $b=0^\circ$, the $x$ axis), along the
Galactic rotation ($l=90^\circ$, $b=0^\circ$, the $y$ axis), and
toward the North Galactic Pole ($b=90^\circ$, the $z$ axis). In
the Ogorodnikov–Milne model, we use the notation introduced by
Clube (1972, 1973) and employed by du Mont (1977, 1978). When
using only the stellar proper motions, one of the diagonal terms
of the deformation matrix is known (Ogorodnikov 1965) to remain
indeterminate. It is possible to determine the differences between
the diagonal elements of the deformation matrix, for example, in
the form ($M_{11}^+ - M_{22}^+)$ and ($M_{33}^+ -M_{22}^+$).
 In this approach, the basic equations can be written as
$$\displaylines{\hfill
 \mu_{l}\cos b=       (1/r)(X_{\odot}\sin l-Y_{\odot}\cos l) 
   -M_{\scriptscriptstyle32}^{\scriptscriptstyle-}\cos l\sin b
   -M_{\scriptscriptstyle13}^{\scriptscriptstyle-}\sin l\sin b
   +M_{\scriptscriptstyle21}^{\scriptscriptstyle-}\cos b+
 \hfill\llap(1)\cr\hfill
   +M_{\scriptscriptstyle12}^{\scriptscriptstyle+}\cos 2l\cos b
   -M_{\scriptscriptstyle13}^{\scriptscriptstyle+}\sin l\sin b
   +M_{\scriptscriptstyle23}^{\scriptscriptstyle+}\cos l\sin b
  -0.5(M_{\scriptscriptstyle11}^{\scriptscriptstyle+}
  -M_{\scriptscriptstyle22}^{\scriptscriptstyle+})\sin 2l\cos b,
\hfill \cr\hfill
\mu_b=
    (1/r)(X_{\odot}\cos l\sin b+Y_{\odot}\sin l\sin b-Z_{\odot}\cos b)
   +M_{\scriptscriptstyle32}^{\scriptscriptstyle-}\sin l
   -M_{\scriptscriptstyle13}^{\scriptscriptstyle-}\cos l-
 \hfill\llap(2) \cr\hfill
-0.5M_{\scriptscriptstyle12}^{\scriptscriptstyle+}\sin 2l\sin 2b
   +M_{\scriptscriptstyle13}^{\scriptscriptstyle+}\cos l\cos 2b
   +M_{\scriptscriptstyle23}^{\scriptscriptstyle+}\sin l\cos 2b-
 \hfill\cr\hfill
-0.5(M_{\scriptscriptstyle11}^{\scriptscriptstyle+}
    -M_{\scriptscriptstyle22}^{\scriptscriptstyle+})\cos^2 l\sin 2b
+0.5(M_{\scriptscriptstyle33}^{\scriptscriptstyle+}
    -M_{\scriptscriptstyle22}^{\scriptscriptstyle+})\sin 2b,
\hfill }
$$
where $X_\odot$, $Y_\odot$, and $Z_\odot$ are the velocity
components of the peculiar solar motion, and $M_{21}^-$,
$M_{13}^-$, and $M_{32}^-$ are the vector components of the rigid
rotation of an infinitesimal solar neighborhood around the
corresponding axes. In accordance with the adopted rectangular
coordinate system, the following rotations are positive: from axis
1 to axis 2, from axis 2 to axis 3, and from axis 3 to axis 1. The
quantity $M_{21}^+$ is an analogue of the Oort constant $B$. Each
of the quantities $M_{12}^+$, $M_{13}^+$, and $M_{23}^+$ describes
the deformation in the corresponding plane. The quantity
$M_{12}^+$ is an analogue of the Oort constant $A$. The diagonal
components of the deformation tensor $M_{11}^+$, $M_{22}^+$, and
$M_{33}^+$ describe the overall contraction or expansion of the
entire stellar system. Equations (1)--(2) contain eleven unknowns
that can be determined by the least-squares method. The quantity
$1/r$ is the parallactic factor, which is assumed to be equal to
unity. In this case, the stars are referred to a unit sphere. In
this approach, all of the parameters being determined are
proportional to the heliocentric distance of the stellar centroid
under consideration and are expressed in the same units as the
stellar proper motion components, i.e., in mas yr$^{-1}$. Before
the appearance of the Hipparcos catalog, researchers were forced
to use this method of analysis because of the lack of highly
accurate stellar parallaxes. When the distances to stars are
known, the parallactic factor is $1/r=\pi/4.74$, where $\pi$ is
the parallax, and the factor 4.74 is equal to the ratio of the
number of kilometers in an astronomical unit to the number of
seconds in a tropical year. To express the solar velocity
components (mas yr$^{-1}$) in km s$^{-1}$, they must be multiplied
by $4.74/\pi$; to express any of the derived (in mas yr$^{-1}$)
components of the deformation and rotation tensors in km s$^{-1}$
kpc$^{-1}$, they must b e multiplied only by the proportionality
factor 4.74.

{\footnotesize
 \begin{table}[p]
 \caption[]
{\small Kinematic parameters inferred from the proper motions of
Hipparcos stars.}
\medskip
{
 \small
\begin{center}
\parbox{12truecm}{%
\offinterlineskip \halign {#\vrule
&#\hfil&\hfil#&\hfil#&\hfil#&\hfil#&\hfil#
&\hfil#&\hfil#&\vrule#&\hfil#&\hfil#& \vrule#\cr \noalign{\hrule}
&&&&&&&&&&&&height3pt\cr &  &  1~& 2~&  3~&  4~&  5~&  6~& 7~& &
8~&&\cr &&&&&&&&&&&&height3pt\cr \noalign{\hrule}
&&&&&&&&&&&&height3pt\cr &~$\overline r,$~kpc  &   $0.074$~&
$0.142$~&   $0.243$~&   $0.361$~& $0.570$~&$1.2$~&$\sim 2$~&&
~~$r>0.2$~&&\cr&&&&&&&&&&&&height6pt\cr &~$N_{\star}$    & 13453~&
29378~&     20032~&     16040~& 11441~& 10015~& 3833~&&
58675~&&\cr&&&&&&&&&&&&height3pt\cr
\noalign{\hrule}&&&&&&&&&&&&height6pt\cr &~$X_{\odot}$    &
$25.32$~&   $14.27$~&    $8.10$~&  $ 6.05$~&
                     $ 4.53$~&    $3.56$~&    $2.23$~&& $ 5.89$~&&\cr&&&&&&&&&&&&height1pt\cr
&                & $\pm0.70$~& $\pm0.25$~& $\pm0.17$~& $\pm0.14$~&
                   $\pm0.13$~& $\pm0.11$~& $\pm0.19$~&&$\pm0.08$~&&\cr&&&&&&&&&&&&height4pt\cr
&~$Y_{\odot}$    &   $47.97$~&   $23.24$~&   $13.37$~&  $10.24$~&
                     $ 8.06$~& $ 6.75$~& $ 5.81$~&& $10.22$~&&\cr&&&&&&&&&&&&height1pt\cr
&~               & $\pm0.69$~& $\pm0.13$~& $\pm0.18$~& $\pm0.15$~&
                   $\pm0.14$~& $\pm0.13$~& $\pm0.18$~&& $\pm0.08$~&&\cr&&&&&&&&&&&&height4pt\cr
&~$Z_{\odot}$    &   $18.59$~&   $10.04$~&   $ 5.66$~&  $ 4.27$~&
                     $ 3.23$~&   $ 2.49$~&   $ 2.03$~&& $4.10$~&&\cr&&&&&&&&&&&&height1pt\cr
&~               & $\pm0.68$~& $\pm0.24$~& $\pm0.16$~& $\pm0.13$~&
                   $\pm0.12$~& $\pm0.10$~& $\pm0.16$~&& $\pm0.07$~&&\cr&&&&&&&&&&&&height8pt\cr
&~$V_{\odot}$    &   $20.1~$~&   $19.6~$~&   $19.1~$~&  $21.6~$~&
                     $26.4~$~&   $45.7~$~&   $\sim 62.0~$~&& $22.0~$~&&\cr&&&&&&&&&&&&height1pt\cr
&~               & $\pm0.2~$~& $\pm0.2~$~& $\pm0.2~$~& $\pm0.2~$~&
                   $\pm0.4~$~& $\pm0.7~$~& $\pm1.7~$~&& $\pm0.1~$~&&\cr&&&&&&&&&&&&height8pt\cr
&~$L_{\odot}$    & $62.2~$~& $58.6~$~&$58.8~$~& $59.4~$~&
$60.6~$~&
                   $62.2~$~& $69.0~$~&& $60.0~$~&&\cr&&&&&&&&&&&&height1pt\cr
&~             & $\pm0.7~$~& $\pm0.5~$~& $\pm0.6~$~& $\pm0.7~$~&
                 $\pm0.8~$~& $\pm0.9~$~& $\pm1.7~$~&&$\pm0.4~$~&&\cr&&&&&&&&&&&&height4pt\cr
&~$B_{\odot}$    & $18.9~$~& $20.2~$~&$19.9~$~& $19.8~$~&
$19.3~$~&
                   $18.0~$~& $18.1~$~&&$19.2~$~&&\cr&&&&&&&&&&&&height1pt\cr
&~               & $\pm0.7~$~& $\pm0.5~$~& $\pm0.6~$~& $\pm0.6~$~&
                   $\pm0.7~$~& $\pm0.9~$~& $\pm1.4~$~&&$\pm0.3~$~&&\cr&&&&&&&&&&&&height8pt\cr
&~$M_{\scriptscriptstyle12}^{\scriptscriptstyle+}$
                 &$ 2.40$~& $ 2.20$~& $ 2.96$~&
                  $ 2.85$~& $ 2.77$~& $ 2.91$~&
                  $ 2.48$~&&$2.90$~&&\cr&&&&&&&&&&&&height1pt\cr
&~
                 &$\pm0.86$~& $\pm0.30$~& $\pm0.21$~&
                  $\pm0.17$~& $\pm0.16$~& $\pm0.14$~&
                  $\pm0.21$~&&$\pm0.09$~&&\cr&&&&&&&&&&&&height8pt\cr
&~$M_{\scriptscriptstyle32}^{\scriptscriptstyle-}$
                 &$-0.18$~& $ 0.06$~& $-0.15$~&
                  $-0.10$~& $ 0.07$~& $-0.11$~& $-0.53$~&& $-0.16$~&&\cr&&&&&&&&&&&&height1pt\cr
&~
                 &$\pm0.70$~& $\pm0.25$~& $\pm0.18$~&
                  $\pm0.15$~& $\pm0.14$~& $\pm0.13$~& $\pm0.21$~&&
                  $\pm0.08$~&&\cr&&&&&&&&&&&&height8pt\cr
&~$M_{\scriptscriptstyle13}^{\scriptscriptstyle-}$
                 &$-2.44$~& $-1.06$~& $-0.62$~&
                  $-0.15$~& $-0.24$~& $-0.18$~& $-0.68$~&& $-0.36$~&&\cr&&&&&&&&&&&&height1pt\cr
&~
                 &$\pm0.69$~& $\pm0.25$~& $\pm0.18$~&
                  $\pm0.15$~& $\pm0.15$~& $\pm0.13$~& $\pm0.21$~&&
                  $\pm0.09$~&&\cr&&&&&&&&&&&&height8pt\cr
&~$M_{\scriptscriptstyle21}^{\scriptscriptstyle-}$
                 &$-2.40$~& $-2.47$~& $-3.01$~&
                  $-3.03$~& $-3.00$~& $-2.85$~& $-2.76$~&& $-2.93$~&&\cr&&&&&&&&&&&&height1pt\cr
&~
                 &$\pm0.68$~&  $\pm0.24$~& $\pm0.16$~&
                  $\pm0.13$~&  $\pm0.12$~& $\pm0.10$~&
                  $\pm0.16$~&& $\pm0.07$~&&\cr&&&&&&&&&&&&height8pt\cr
&~$M_{\scriptscriptstyle11}^{\scriptscriptstyle+}-M_{\scriptscriptstyle22}^{\scriptscriptstyle+}$
&
                  $ 0.08$~& $-2.13$~& $-2.00$~&$-1.40$~& $-1.16$~& $-1.50$~& $-0.62$~&& $-1.35$~&&\cr&&&&&&&&&&&&height1pt\cr
&~               &$\pm1.76$~&  $\pm0.62$~& $\pm0.42$~& $\pm0.34$~&
$\pm0.32$~& $\pm0.27$~&
                  $\pm0.41$~&& $\pm0.19$~&&\cr&&&&&&&&&&&&height8pt\cr
&~$M_{\scriptscriptstyle13}^{\scriptscriptstyle+}$
                 &$-4.01$~& $-0.56$~& $-0.10$~&
                  $-0.12$~& $-0.14$~& $-0.07$~& $-0.75$~&& $-0.11$~&&\cr&&&&&&&&&&&&height1pt\cr
&~
                 &$\pm0.92$~&  $\pm0.33$~& $\pm0.23$~&
                  $\pm0.20$~&  $\pm0.19$~& $\pm0.17$~&
                  $\pm0.25$~&& $\pm0.11$~&&\cr&&&&&&&&&&&&height8pt\cr
&~$M_{\scriptscriptstyle23}^{\scriptscriptstyle+}$
                 &$-1.38$~& $-0.58$~& $ 0.41$~&
                  $-0.03$~& $-0.20$~& $ 0.14$~& $ 0.49$~&& $0.13$~&&\cr&&&&&&&&&&&&height1pt\cr
&~
                 &$\pm0.92$~&  $\pm0.33$~& $\pm0.23$~&
                  $\pm0.19$~&  $\pm0.18$~& $\pm0.16$~&
                  $\pm0.25$~&& $\pm0.10$~&&\cr&&&&&&&&&&&&height8pt\cr
&~$M_{\scriptscriptstyle33}^{\scriptscriptstyle+}-M_{\scriptscriptstyle22}^{\scriptscriptstyle+}$
&
                  $-1.65$~& $-1.33$~& $-0.61$~& $ 0.17$~&
                  $-0.07$~& $-0.24$~& $-0.39$~&& $-0.14$~&&\cr&&&&&&&&&&&&height1pt\cr
&~                &$\pm1.73$~& $\pm0.63$~& $\pm0.44$~& $\pm0.37$~&
                  $\pm0.36$~& $\pm0.33$~& $\pm0.52$~&& $\pm0.21$~&&\cr&&&&&&&&&&&&height8pt\cr
&~$l_{xy}$  &$-1\pm11$~& $13\pm4~$~& $ 9\pm2~$~& $ 7\pm2~$~&
             $ 6\pm2~$~& $ 7\pm1~$~& $ 4\pm2~$~&&$ 7\pm1~$~&&\cr&&&&&&&&&&&&height4pt\cr
\noalign{\hrule}}}%
 \end{center}}
{\small Note: ${\overline r}$ is the mean distance, $N_\star$ is
the number of stars, the coordinates of the solar apex $L_\odot$
and $B_\odot$ and the vertex $l_{xy}$ are in degrees, $V_\odot$ is
the solar velocity in km s$^{-1}$; all of the remaining quantities
are in mas yr$^{-1}$.}
 \end{table}}

 \bigskip
 \leftline{\hskip6mm THE SYSTEM OF THE HIPPARCOS CATALOG}
 \bigskip
We divided the Hipparcos stars into seven groups, depending on the
heliocentric distance (kpc):
 $0.05-0.1$, $0.1-0.2$, $0.2-0.3$, $0.3-0.45$, $0.45-0.66$, $>0.66$,
and the group of stars with negative parallaxes. We imposed a
constraint on the stellar space velocity that allows us to discard
only those stars whose space velocities exceed the hyperbolic
velocity, for example, stars with enormous peculiar velocities
acquired through explosions and encounters,
 $V=220+62=282$~km s$^{-1}$.
Here, 220 km s$^{-1}$ is the circular velocity of Galactic
rotation at the solar Galactocentric distance (recommended by IAU
1986) and 62 km s$^{-1}$ is the mean stellar space velocity
dispersion in the solar neighborhood (estimated by Oort). The
Hipparcos stars whose space velocities exceed the hyperbolic
velocity were found, for example, by Moffat et al. (1998). In this
case, when using only the stellar proper motions, we estimated the
stellar space velocities on the basis of the Kleiber theorem
(Agekyan et al. 1962):
 $$\displaylines
 {\hfill |V_t|\cdot 4.74\cdot r<{3.14\over 4}\cdot |V|.\hfill}
$$
Here, $V_t=\sqrt{\mu_{l \cos b}^2+\mu_{b}^2}$ is the tangential
velocity of the star in mas yr$^{-1}$, and $V$ is the total space
velocity of the star in km s$^{-1}$. Table 1 presents the results
of the simultaneous solution of Eqs. (1) and (2). Columns 1--7
give the parameters obtained from the proper motions of the stars
located in seven ``thin'' spherical shells. All of the stars with
negative parallaxes were assumed to be located at a heliocentric
distance of 2 kpc. The last column gives the kinematic parameters
determined from the proper motions of 58675 distant stars located
farther than 0.2 kpc from the Sun (the mean heliocentric distance
of this group of stars is 0.371 kpc). We emphasize that these are
special solutions; they were obtained by setting the parallactic
factor equal to $1/r=1$. We needed the parallaxes to divide the
stars into distance-limited groups. This method made it possible,
first, to use the most distant Hipparcos stars with negative
parallaxes, second, to completely eliminate all of the effects of
random parallax errors when solving Eqs. (1) and (2), and, third,
to ensure that the results obtained by this method are in close
agreement with those obtained from the proper motions of TRC
stars. Since we used this approach to control the inertiality of
the Hipparcos catalog, the solution from the last column of Table
1 that is of greatest interest in this respect may be called
inertial. Let us consider the derived kinematic parameters of the
inertial solution (in mas yr$^{-1}$):
$$
\displaylines{\hfill
M_{\scriptscriptstyle12}^{\scriptscriptstyle+}=+2.90\pm0.09,\hfill\cr\hfill
M_{\scriptscriptstyle21}^{\scriptscriptstyle-}=-2.93\pm0.07,\hfill\cr\hfill
M_{\scriptscriptstyle13}^{\scriptscriptstyle+}=-0.11\pm0.11,\hfill\cr\hfill
M_{\scriptscriptstyle13}^{\scriptscriptstyle-}=-0.36\pm0.09,\hfill\llap(3)\cr\hfill
M_{\scriptscriptstyle23}^{\scriptscriptstyle+}=+0.13\pm0.10,\hfill\cr\hfill
M_{\scriptscriptstyle32}^{\scriptscriptstyle-}=-0.16\pm0.08,\hfill\cr\hfill
M_{\scriptscriptstyle11}^{\scriptscriptstyle+}-
M_{\scriptscriptstyle22}^{\scriptscriptstyle+}=-1.35\pm0.19,\hfill\cr\hfill
M_{\scriptscriptstyle33}^{\scriptscriptstyle+}-
M_{\scriptscriptstyle22}^{\scriptscriptstyle+}=-0.14\pm0.21.\hfill
}
$$
 As we can see, the
following quantities are statistically significant:
 $M_{12}^+$,
 $M_{21}^-$, and
 $(M_{11}^+ - M_{22}^+)$.
Based on these quantities, we find that
 $A= 13.7\pm0.4$ km s$^{-1}$ kpc$^{-1}$,
 $B=-13.9\pm0.3$ km s$^{-1}$ kpc$^{-1}$,
 $C= -3.2\pm0.5$ km s$^{-1}$ kpc$^{-1}$ and the vertex
deviation $l_{xy}=7\pm1^\circ$. To this end, we used the standard
relations
 $A= M_{12}^+ \cdot 4.74$,
 $B= M_{21}^- \cdot 4.74$,
 $C=4.74 \cdot (M_{11}^+ -M_{22}^+)/2$
 and $\tan 2l_{xy}=-C/A$.

 \begin{figure} {\begin{center}
 \includegraphics[width=80mm]{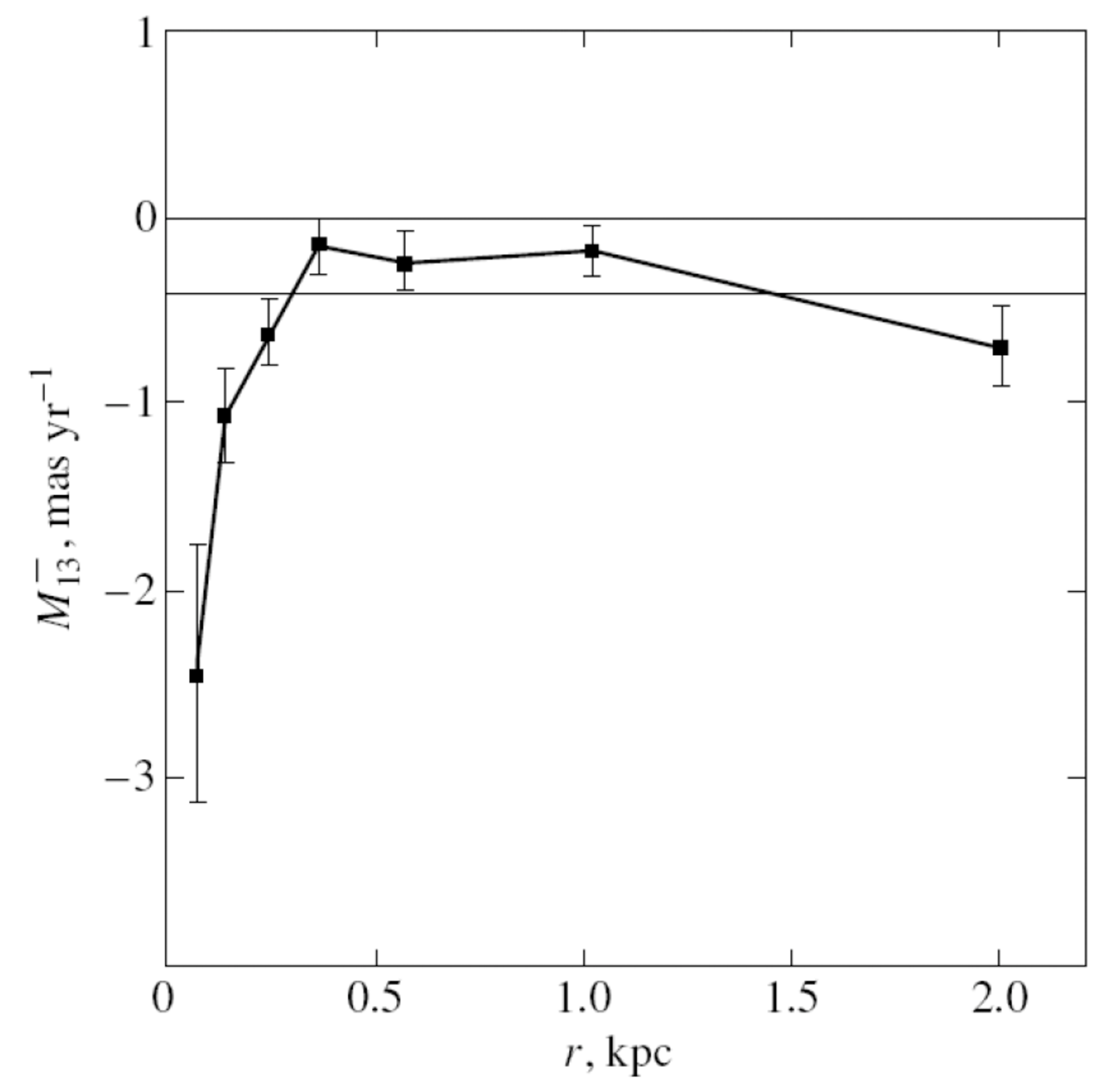}
 \caption{Component $M_{13}^-$ of the rigid-rotation tensor for
Hipparcos stars against stellar heliocentric distance $r$.}
 \label{f1} \end{center} } \end{figure}

The Oort constants $A$ and $B$ obtained by this method are in good
agreement with the values recommended by the IAU (1986) and with
the values from the paper by Bobylev (2004):
 $A= 13.7\pm0.6$ km s$^{-1}$ kpc$^{-1}$,
 $B=-12.9\pm0.4$ km s$^{-1}$ kpc$^{-1}$ in
which they were calculated by analyzing their recent
determinations by various authors. The vertex deviation also
agrees with its published determinations based on various
observational data, for example, with those by Dehnen and Binney
(1998). Note, nevertheless, that $M_{13}^-$ is also statistically
significant. Figure 1 shows a plot of $M_{13}^-$ against
heliocentric distance constructed from the data of Table~1. As we
can see from Table~1 and Fig.~1, $M_{13}^-$ is nonzero for the
most distant Hipparcos stars. It should also be emphasized that,
as can be seen from Fig.~1, the constraint $r>0.2$~kpc used to
obtain the inertial solution and to determine $M_{13}^-$ has such
an effect that we obtain a lower limit on $M_{13}^-$.

 \begin{figure} {\begin{center}
 \includegraphics[width=160mm]{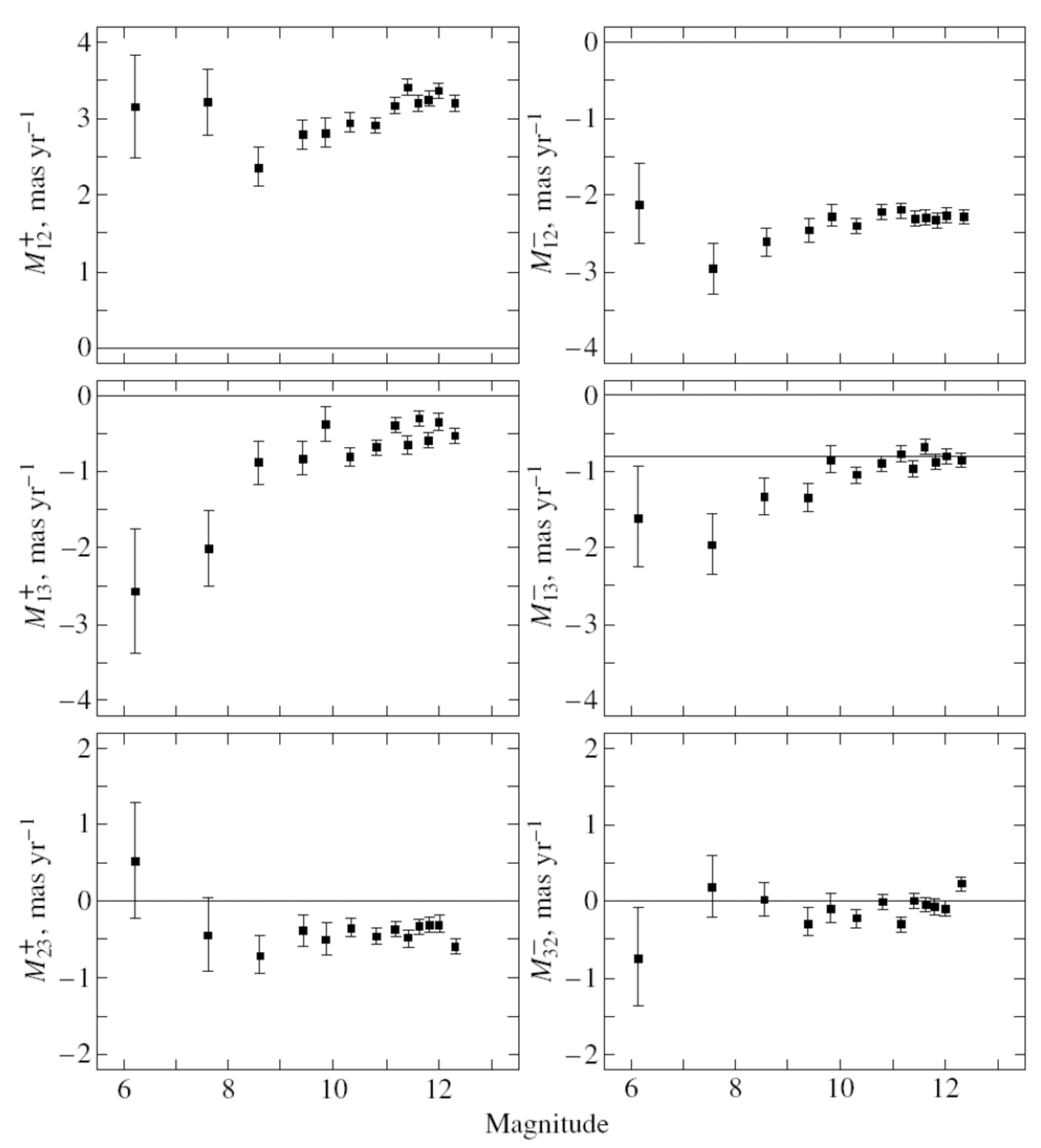}
 \caption{Kinematic parameters inferred from the proper motions of
TRC stars versus magnitude (Tycho B mag.).}
 \label{f2} \end{center} } \end{figure}

 \newpage
\bigskip
\leftline{\hskip6mm THE SYSTEM OF THE TRC CATALOG}
\bigskip
The results of our solution of Eqs. (1) and (2) based on TRC stars
of mixed spectral composition are plotted against magnitude in
Fig.~2. The random errors of all of the sought unknowns are small
in all of the magnitude intervals, each of which contains
$\approx100000$ stars, and are equal to 0.1--0.2 mas yr$^{-1}$ for
faint stars. As we can see from Fig.~2, the parameters that
describe the deformation in the $yz$ plane and the rotation around
the $x$ axis, i.e., $M_{23}^+$ and $M_{32}^-$, are virtually equal
to zero. The parameters that describe the deformations in the $xy$
and $yz$ planes and the rotation around the $z$ and $y$ axes
differ significantly from zero. The values of $M_{13}^-$ are
nonzero for the faintest TRC stars. We calculated the mean,
$$
\displaylines{\hfill
M_{\scriptscriptstyle13}^{\scriptscriptstyle-}=
-0.86\pm0.11~{\hbox {mas yr$^{-1}$}}, \hfill \llap(4) }
$$
by using the results obtained in nine intervals of magnitudes
fainter than $9^m.5$. This result must be compared with the
results of our analysis of the proper motions for Hipparcos stars.
The deviation of the vertex $l_{xy}$ is virtually equal to zero in
each magnitude interval, which does not confirm the conclusions
drawn from our analysis of the Hipparcos stellar proper motions.
This result probably stems from the fact that to compile the
reference catalog that the TRC was intended to be (Kuzmin et al.
1999), the stars were selected kinematically; no high
proper-motion stars were used. We calculated the mean values of
 $M_{21}^+ = 3.15\pm0.20$ mas yr$^{-1}$ and
 $M_{12}^- =-2.27\pm0.06$ mas yr$^{-1}$ by the mean of the
nine values obtained in the intervals of magnitudes fainter than
$9^m.5$ (the mean being $m=11^m.2$). This yields the following
Oort constants:
 $A= 14.93\pm0.97$ km s$^{-1}$ kpc$^{-1}$ and
 $B=-10.77\pm0.31$ km s$^{-1}$ kpc$^{-1}$,
which are in good agreement with those calculated by Olling and
Dehnen (1999) from the Tycho/ACT proper motions of red giants:
 $A= 14.2\pm1$ km s$^{-1}$ kpc$^{-1}$ and
 $B=-12.7\pm1$ km s$^{-1}$ kpc$^{-1}$.
The model component that describes the rotation about the Galactic
$y$ axis inferred here from a kinematic analysis of the proper
motions of TRC stars is $M_y=-0.86\pm0.11$ mas yr$^{-1}$ and
confirms the rotation that we found from the Hipparcos catalog
(solution (3)). In this case, we have an upper limit, since the
data for the faintest stars include the effect of the real
rotation of nearby stars that to the Local stellar system
(Fig.~1).

{\footnotesize\begin{table}[t] \caption[]
 {Precession corrections $\Delta p_1$ and $\Delta E$, mas yr$^{-1}$.}
\begin{center}\begin{tabular}{|r|r|rr|}\hline
                    Reference~&      Data~& $ \Delta p_1$ ~~~~& $\Delta E$ ~~~~\\\hline
    ~~Miyamoto, S\^oma (1994) &      ACRS~& $-2.7_{(0.3)}$  ~~&                \\
          ~~Walter, Ma (1994) &      VLBI~& $-3.6_{(1.1)}$  ~~&                \\
~~Charlot {\it et al.} (1995) &  VLBI+LLR~& $-3.0_{(0.2)}$  ~~&                \\
               ~~Rybka (1995) &       PPM~& $-3.1_{(0.2)}$  ~~& $-1.3_{(0.2)}$ \\
             ~~Bobylev (1997) & PUL2--PPM~& $-2.8_{(0.8)}$  ~~&                \\
     ~~Ma {\it et al.} (1998) &      VLBI~& $-2.84_{(0.04)}$  &                \\
            ~~Vityazev (1999) &  CGC--HIP~& $-3.4_{(1.0)}$  ~~& $-3.3_{(1.0)}$ \\\hline
\end{tabular}\end{center}\end{table}}

\bigskip
 \leftline{\hskip6mm CORRECTION TO THE IAU (1976) PRECESSION CONSTANT}
\bigskip
The proper motions of the Hipparcos stars are free from the effect
of precession of the Earth's axis. However, the Hipparcos catalog
is an extension to the optical range of the ICRF system (Ma et al.
1998), which is based on ground-based VLBI observations of radio
sources. To determine the precession parameters from the analysis
of $M_y$, we use two equations that, given the numerical values
for epoch J2000.0 (Perryman et al. 1997), can be written as
$$
\displaylines{\hfill
 M_x=\Omega_x-0.0965\Delta p_1+0.4838\Delta E,\hfill\llap(5)\cr\hfill
 M_y=\Omega_y+0.8623\Delta p_1-0.7470\Delta E.\hfill\llap(6)
}
$$
Here, $\Omega_x$ and  $\Omega_y$ are the vector components of the
real rigid rotation of the stellar system, and this rotation
actually exists at heliocentric distances up to $r\approx0.2$ kpc.
Assuming that at large distances, $M_x-\Omega_x=0$  and
$M_y-\Omega_y=-0.36\pm0.09$ mas yr$^{-1}$ includes only the
precession quantities, we obtain from the solution of Eqs. (5) and
(6)
$$
\displaylines{\hfill
 \Delta p_1=-0.50\pm0.13~{\hbox {mas yr$^{-1}$}},\hfill\llap(7)\cr\hfill
 \Delta   E=-0.10\pm0.02~{\hbox {mas yr$^{-1}$}}.\hfill\llap(8)
 }
$$
Setting $\Delta E=0$, we obtain the following estimate from Eq.
(6):
$$
\displaylines{\hfill \Delta p_1=-0.42\pm0.10~{\hbox {mas
yr$^{-1}$}}.\hfill\llap(9) }
$$
Table 2 gives the results of the determinations of two precession
parameters by various authors: the correction to the IAU (1976)
constant of lunisolar precession in longitude, $\Delta p_1$, and
$\Delta E$, which is the sum of the corrections to the rate of
planetary precession and the motion of the zero point of right
ascensions. Miyamoto and S\^oma (1994) determined the precession
correction based on a kinematic analysis of the proper motions of
ACRS stars using the Ogorodnikov–Milne kinematic model. Walter and
Ma (1994) determined the precession correction based on VLBI
observations of extragalactic radio sources from an annual
catalog. Charlot et al. (1995) determined the precession
correction by analyzing a 24-yr-long series of lunar laser radar
observations and a 16-yr-long series of radiointerferometric
observations. Rybka (1995) determined the precession parameters
based on a kinematic analysis of the proper motions of PPM stars
using the Ogorodnikov–Milne kinematic model. Bobylev (1997)
determined the precession correction by analyzing the differences
between the Pulkovo absolute proper motions and PPM. To obtain the
ICRF system that is based on ground-based VLBI observations of
radio sources, Ma et al. (1998) adopted the following correction
to the IAU (1976) lunisolar precession constant:
 $\Delta p_1 = -2.84\pm0.04$ mas yr$^{-1}$.
Vityazev (1999) determined the precession parameters by comparing
the proper motions of CGC and Hipparcos stars. Our kinematic
analysis of the proper motions of Hipparcos stars yielded $\Delta
p_1 =-0.42\pm0.10$ mas yr$^{-1}$ (solution (9)). It may be assumed
that Ma et al. (1998) slightly underestimated $\Delta p_1$. Since
the ICRS was constructed using precisely this value, we obtain a
statistically significant ``addition'' to this correction, $\Delta
p_1$, when analyzing the Hipparcos and TRC catalogs. Therefore,
the correction to the IAU (1976) lunisolar precession constant is
$$
\displaylines
 {\hfill \Delta p_1=-3.26\pm0.10~{\hbox {mas yr$^{-1}$}}.\hfill\llap(10)}
$$
This value agrees well with the data given in Table 2.

 \begin{figure} {\begin{center}
 \includegraphics[width=80mm]{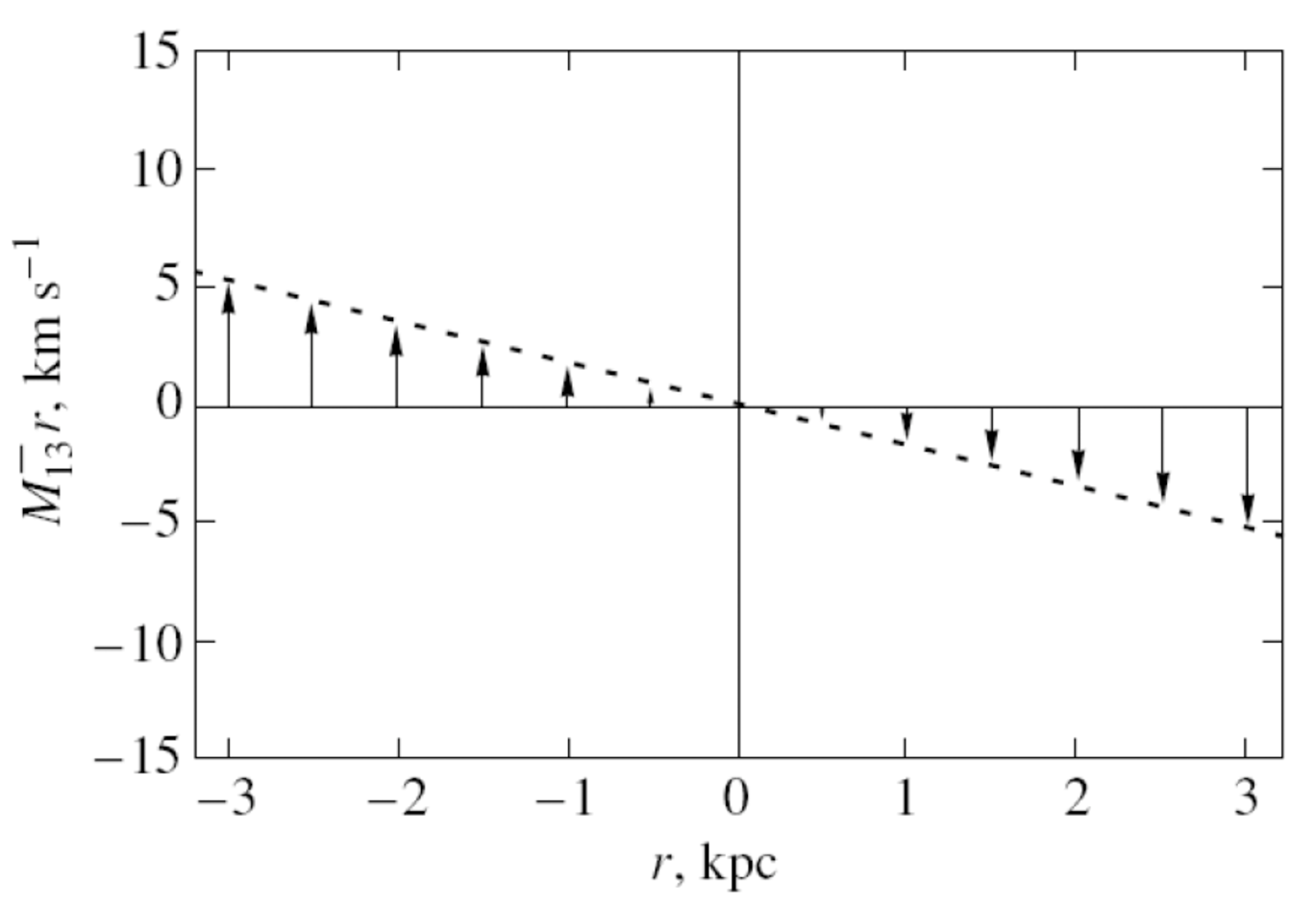}
 \caption{Linear velocity vectors $M_{13}^- \cdot r=W$ versus
heliocentric distance $r.$ The Galactic plane coincides with the
$xz$ plane, the $x$ axis coincides with the $r$ direction, and the
$z$ axis is directed upward.}
 \label{f3} \end{center} } \end{figure}
 \begin{figure} {\begin{center}
 \includegraphics[width=80mm]{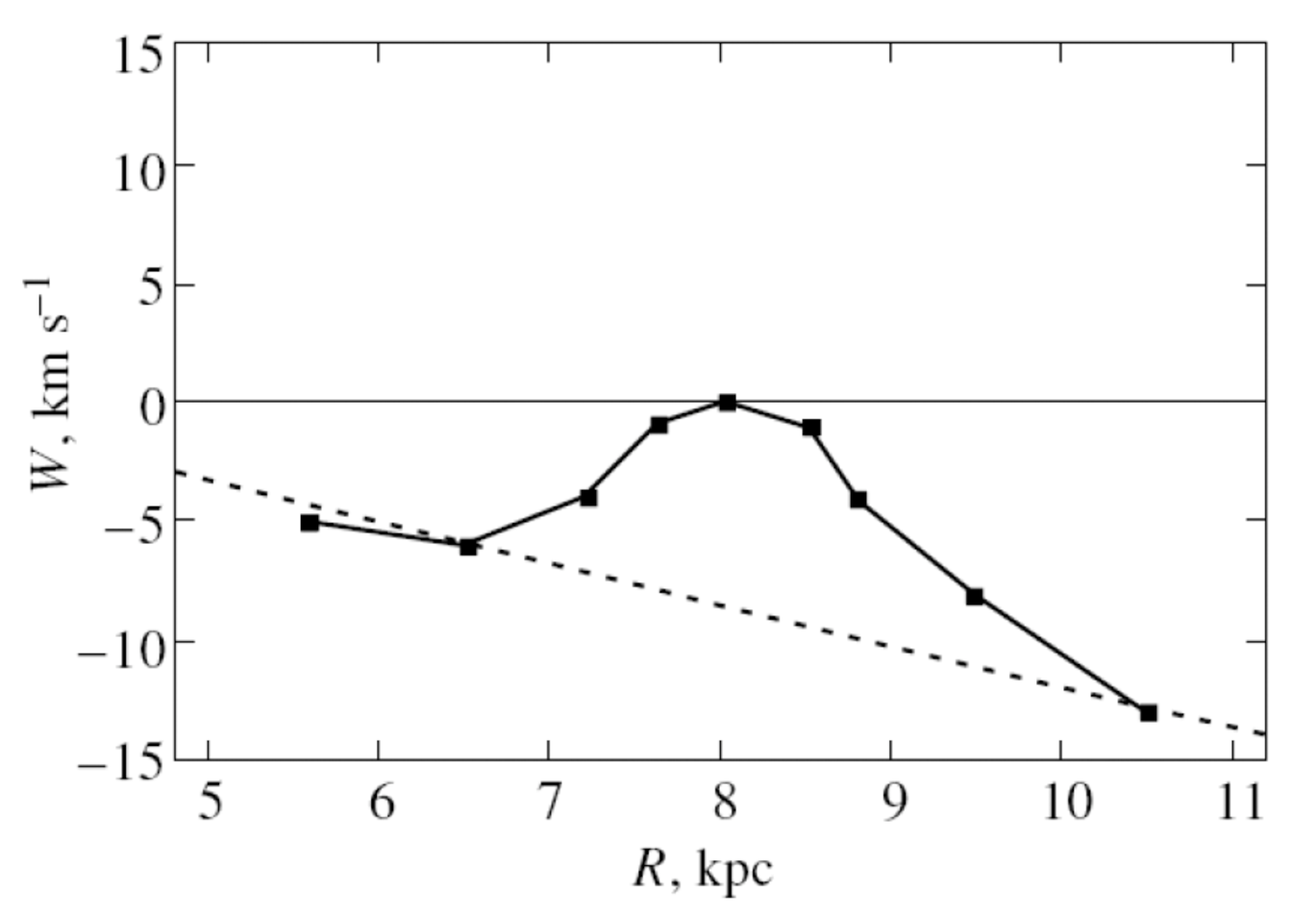}
 \caption{Stellar residual velocity components $W$ versus heliocentric
distance $R.$ The solar motion with respect to the LSR with
velocities $(U,V,W)_\odot =(9,5,7)$ km s$^{-1}$ was taken into
account. The dotted line indicates our dependence, which must be
eliminated.}
 \label{f4} \end{center} } \end{figure}

\bigskip
\leftline{\hskip6mm THE MOTION OF DISTANT OB STARS}
\bigskip
Results (7)--(9) have yet another important implication: they show
that $M_y=-0.36\pm0.09$ mas yr$^{-1}$ may have a nature unrelated
to the real stellar motions. The HI layer in the Galaxy is known
to have an extended warp (Kulikovskii 1985; Carroll and Ostly
1996). Smart et al. (1997) and Drimmel et al. (2000) considered
the hypothesis that the extended Galactic HI layer affects the
kinematics of OB stars. Drimmel et al. (2000) used simulations to
estimate the angular velocity of the precession of OB stars in the
$zx$ plane, $-25$~km s$^{-1}$ kpc$^{-1}$. The effect of our
 $M_{13}^-=-0.36$ mas yr$^{-1}$
(solution (3)) on the kinematic parameters of stars is easy to
calculate. Since
 $M_{13}^-\cdot 4.74=-1.7$ km s$^{-1}$ kpc$^{-1}$, it is equal to
 $-1.7$ and $-3.4$ km s$^{-1}$ at $r=1$ and 2 kpc, respectively. The
presence of such a fictitious wave could completely explain the
precession velocity of $-25$~km s$^{-1}$ kpc$^{-1}$ found by
Drimmel et al. (2000) for distant OB stars. In Fig.~3, the linear
rotation velocity vectors $M_{13}^-r$ are plotted against
heliocentric distance $r$. This rotation is attributable to the
 $M_{13}^- =-0.36$ mas yr$^{-1}$
that we inferred from the Hipparcos catalog. The directions of the
vectors are indicated in accordance with the fact that the
rotation takes place in the $zx$ plane and that, according to the
chosen coordinate system, the rotation from the $z$ axis to the
x-axis is positive. In Fig.~4, the W components of space velocity
(the linear velocity along the z-coordinate) are plotted against
heliocentric distance R. The points are plotted by using the data
obtained by Drimmel et al. (2000) for a sample of 4250 Hipparcos
OB stars in the magnitude interval 0–13$^m$. A comparison of
Fig.~3 (the Sun is assumed to be at a Galactocentric distance of
about $R_\circ=8.0$~kpc) and Fig.~4 leads us to conclude that the
value of $M_{13}^- =-0.36$ mas yr$^{-1}$ that we inferred from
Hipparcos data almost completely explains the slope of the plot in
Fig.~4. Eliminating the dependence that we found should
significantly reduce the precession of $-25$ km s$^{-1}$
kpc$^{-1}$ for distant OB stars derived by Drimmel et al. (2000).
In this case, the fact that distant OB stars belong to the
hydrogen layer implies that there is only a linear displacement of
all distant OB stars along the $z$ coordinate.

\bigskip
\leftline {\hskip6mm CONCLUSIONS}
\bigskip
Based on a linear Ogorodnikov-Milne model, we have performed a
kinematic analysis of the Hipparcos and TRC stellar proper
motions. We found the rotation of all distant ($r>0.2$~kpc)
Hipparcos stars with a mean angular velocity of
 $M_{13}^-=-0.36\pm0.09$ mas yr$^{-1}$ about the Galactic $y$ axis. One of the
causes of this rotation may be an inaccuracy of the lunisolar
precession constant adopted when creating the ICRF (Ma et al.
1998). We showed that, in this case, the correction to the IAU
(1976) constant of lunisolar precession in longitude is
 $\Delta p_1=-3.26\pm0.10$ mas yr$^{-1}$.
We have shown that eliminating the rotation
$M_{13}^-=-0.36\pm0.09$ mas yr$^{-1}$ from the proper motions of
distant OB stars should lead to a significant reduction in the
$-25$~km s$^{-1}$ kpc$^{-1}$ precession of distant OB stars
inferred by Drimmel et al. (2000). In this case, the fact that
distant OB stars belong to the hydrogen layer reduces only to a
linear displacement of all distant OB stars along the $z$
coordinate. We have determined the mean Oort constants
 $A= 14.9\pm1.0$ km s$^{-1}$ kpc$^{-1}$ and
 $B=-10.8\pm0.3$ km s$^{-1}$ kpc$^{-1}$ from TRC data. The model
component that describes the rotation of all TRC stars about the
Galactic $y$ axis was found to be nonzero for all magnitudes,
$M_{13}^-=-0.86\pm0.11$ mas yr$^{-1}$.

 \bigskip
 \leftline {\hskip6mm ACKNOWLEDGMENTS}
 \medskip
This work was supported by the Russian Foundation for Basic
Research (project no. 02--02--16570).

\bigskip
\leftline {\hskip6mm REFERENCES}
 {\small

1.~T.A.~Agekyan, B.A.~Vorontsov-Vel’yaminov, V.G.~Gorbatskii, et
al., {\it A Course on Astrophysics and Stellar Astronomy, Ed. by
A.A. Mikhailov} (Fizmatlit, Moscow, 1962), Vol. 2 [in Russian].

2.~V.V.~Bobylev, {\it JOURNEES 1997}, Ed. by J. Vondrak and N.
Capitaine (Obs. de Paris, Paris, 1997), p. 91.

3.~V.V.~Bobylev, Pis’ma Astron. Zh. {\bf 30}, 185 (2004).

4.~B.W.~Carroll and D.A.~Ostly, {\it An Introduction to Modern
Astrophysics} (Addison-Wesley, New York, 1996).

5.~P.~Charlot, O.J.~Sovers, J.G.~Williams, {\it et al.}, Astron.
J. {\bf 109}, 418 (1995).

6.~S.V.M. Clube, Mon. Not. R. Astron. Soc. {\bf 159}, 289 (1972).

7.~S.V.M. Clube, Mon. Not. R. Astron. Soc. {\bf 161}, 445 (1973).

8.~W.~Dehnen and J.J.~Binney, Mon. Not. R. Astron. Soc. {\bf 298},
387 (1998).

9.~R.Drimmel, R.L. Smart, and M.G.~Lattanzi, Astron. Astrophys.
{\bf 354}, 67 (2000).

10.~E. H\o g, A.~Kuzmin, U.~Bastian, et al., Astron.Astrophys.
333, L65 (1998).

11.~A.~Kuzmin, E. H\o g, U. Bastian, et al., Astron.Astrophys.,
Suppl. Ser. {\bf 136}, 491 (1999).

12.~P.G. Kulikovskii, {\it Stellar Astronomy} (Nauka, Moscow,
1985) [in Russian].

13.~C.~Ma, E.F.~Arias, T.M.~Eubanks, et al., Astron. J. {\bf 116},
516 (1998).

14.~M.~Miyamoto and M.~S\^oma, Astron. J. {\bf 105}, 691 (1993).

15.~B. du Mont, Astron. Astrophys. {\bf 61}, 127 (1977).

16.~B. du Mont, Astron. Astrophys. {\bf 66}, 441 (1978).

17.~A.F.G. Moffat, S.V. Marchenko, W. Seggewiss, {\it et al.},
Astron. Astrophys. {\bf 331}, 949 (1998).

18.~K.F. Ogorodnikov, {\it Dynamics of Stellar Systems}
(Fizmatgiz, Moscow, 1965).

19.~R.P.~Olling and W. Dehnen, Bull. Am. Astron. Soc. {\bf 31},
1379 (1999).

20. M.A.C. Perriman, L. Lindegren, J. Kovalevsky, et al., Eur.
Space Agency 1–17 (1997).

21.~S.P.~Rybka, Kinemat. Fiz. Neb. Tel {\bf 11}, 77 (1995).

22.~R.L.~Smart, R. Drimmel, M.G. Lattanzi, et al.,
 {\it JOURNEES 1997}, Ed. by J. Vondrak and N. Capitaine (Obs. de Paris, Paris,
1997), p. 179.

23. The HIPPARCOS and Tycho Catalogues, ESA SP-1200 (1997).

24.~V.V. Vityazev, Doctoral Dissertation in Mathematical Physics
(St. Petersburg, 1999) [in Russian].

25.~H.G.~Walter and C. Ma, Astron. Astrophys. {\bf 284}, 1000
(1994).
 }

\end{document}